\documentclass[fleqn,11pt,twoside]{article}

\usepackage{amsthm,amsthm,amssymb, color, xcolor,epsfig, graphics, subfigure,comment}

\usepackage{amsmath, graphicx, latexsym, lscape}
\usepackage{float}

\makeatletter
\newcommand{\copyrightnote}[2]{{\renewcommand{\thefootnote}{}
 \footnotetext{\small\it
\begin{flushleft}
 \copyright \ #1   #2
\end{flushleft}}}}

\newcommand{\Name}[1]{\begin{flushleft}
                       \LARGE \bf #1
                       \end{flushleft}\vspace{-3mm}}

\newcommand{\Author}[1]{\begin{flushleft}
                       \it #1 \end{flushleft}}

\newcommand{\Address}[1]{\begin{flushleft}
                       \it #1 \end{flushleft}}

\newcommand{\Date}[1]{\begin{flushleft}
                      \small  \it #1 \end{flushleft}}

\newcommand{\evenhead}{Author \ name}
\newcommand{\oddhead}{Article \ name}

\renewcommand{\@evenhead}{
\hspace*{-3pt}\raisebox{-15pt}[\headheight][0pt]{\vbox{\hbox to \textwidth
{\thepage \hfil \evenhead}\vskip4pt \hrule}}}
\renewcommand{\@oddhead}{
\hspace*{-3pt}\raisebox{-15pt}[\headheight][0pt]{\vbox{\hbox to \textwidth
{\oddhead \hfil \thepage}\vskip4pt\hrule}}}
\renewcommand{\@evenfoot}{}
\renewcommand{\@oddfoot}{}

\long\def\@makecaption#1#2{%
  \vskip\abovecaptionskip
  \sbox\@tempboxa{\small {\bf #1.}\ \ #2}%
  \ifdim \wd\@tempboxa >\hsize
    {\small {\bf #1.}\ \ #2}\par
  \else
    \global \@minipagefalse
    \hb@xt@\hsize{\hfil\box\@tempboxa\hfil}%
  \fi
  \vskip\belowcaptionskip}

\newcommand{\JNMPnumberwithin}[3][\arabic]{%
  \@ifundefined{c@#2}{\@nocounterr{#2}}{%
    \@ifundefined{c@#3}{\@nocnterr{#3}}{%
      \@addtoreset{#2}{#3}%
      \@xp\xdef\csname the#2\endcsname{%
        \@xp\@nx\csname the#3\endcsname .\@nx#1{#2}}}}%
}

\newcommand{\resetfootnoterule} {
  \renewcommand\footnoterule{%
  \kern-3\p@
  \hrule\@width.4\columnwidth
  \kern2.6\p@}
}

\renewcommand{\footnoterule}{}

\makeatother

\theoremstyle{definition}

\begin{document}

\renewcommand{\evenhead}{ {\LARGE\textcolor{blue!10!black!40!green}{{\sf \ \ \ ]ocnmp[}}}\strut\hfill P Albares, P G Est\'evez, A Gonz\'alez-Parra and P del Olmo}
\renewcommand{\oddhead}{ {\LARGE\textcolor{blue!10!black!40!green}{{\sf ]ocnmp[}}}\ \ \ \ \  Spectral problem for the complex mKdV equation}

\thispagestyle{empty}
\newcommand{\FistPageHead}[3]{
\begin{flushleft}
\raisebox{8mm}[0pt][0pt]
{\footnotesize \sf
\parbox{150mm}{{Open Communications in Nonlinear Mathematical Physics}\ \ \ \ {\LARGE\textcolor{blue!10!black!40!green}{]ocnmp[}}
\quad Special Issue 1, 2024\ \  pp
#2\hfill {\sc #3}}}\vspace{-13mm}
\end{flushleft}}

\FistPageHead{1}{\pageref{firstpage}--\pageref{lastpage}}{ \ \ }

\strut\hfill

\strut\hfill

\copyrightnote{The author(s). Distributed under a Creative Commons Attribution 4.0 International License}

\begin{center}
{  {\bf This article is part of an OCNMP Special Issue\\
\smallskip
in Memory of Professor Decio Levi}}
\end{center}

\smallskip

\Name{Spectral problem for the complex mKdV equation: singular  manifold method and Lie symmetries}

\Author{P. Albares$^{\,1,2}$,  P. G. Est\'evez$^{\,1}$, A. González-Parra$^{1}$, P.  del Olmo$^{,\,1}$}

\Address{$^{1}$ Departamento de F\'isica Fundamental, Universidad de Salamanca, 37008 Salamanca, Spain \\[2mm]
$^{2}$ Departament of Mathematics, Heriot-Watt University, Edinburgh EH14 4AS, United Kingdom}

\Date{Received July 22, 2023; Accepted October 3, 2023}

\setcounter{equation}{0}

\begin{abstract}

\noindent
This article addresses the study of the complex version of the modified Korteweg-de Vries equation using two different approaches. Firstly, the singular manifold method is applied in order to obtain the associated spectral problem, binary Darboux transformations and $\tau$-functions. The second part concerns the identification of the classical Lie symmetries for the spectral problem. The similarity reductions associated with these symmetries allow us to derive the reduced spectral problems and first integrals for the ordinary differential equations arising from such reductions.

\end{abstract}

\label{firstpage}


\section{Introduction}
\label{sec:intro}


The integrability analysis of nonlinear partial differential equations (PDEs) can be addressed from different perspectives, which are not always shown to be equivalent \cite{ecg98}. The existence of a Lax pair \cite{lax1968} that permits to solve the associated spectral problem in terms of the inverse scattering transform (IST) \cite{ac91,ablowitz78} is frequently considered as a definite proof of the integrability of a given PDE. Nevertheless, the determination of the associated linear problem is neither trivial nor straightforward task. In most cases, the aforementioned spectral problem is obtained by inspection. This quest becomes even harder when dealing with nonlinear ordinary differential equations (ODEs). Lax pairs for ODEs involve derivatives with respect to the spectral parameter and it is exceedingly difficult to compute them by inspection \cite{levi87}, especially for nonautonomous equations.

In this paper we  focus on two of these possible approaches which, in our view, aid in determining both the spectral problem associated with a nonlinear PDE and the Lax pair of an ODE arising from a reduction of a PDE whose spectral problem is known:

\begin{itemize}

\item{\bf Painlevé Property}

The first approach is based on the Painlevé Property \cite{painleve} and the conjecture that if a differential equation successfully passes the Painlevé test, it should be possible to find an associated spectral problem (Lax pair) \cite{lax1968}. Within this framework, the determination of the spectral problem is frequently achieved using the singular manifold method (SMM) \cite{weiss}. While the SMM provides the suitable results in most cases, it may present some limitations, especially when the equation has multiple Painlevé branches \cite{egmm} or if it has resonances in the dominant terms \cite{albares2019}. Several modifications of the SMM that allow us to overcome these issues can be found in \cite{ecg98}.

\item{\bf Lie symmetries}

Lie symmetries constitute a powerful technique to tackle any sort of PDEs, regardless of their integrability. Of capital importance has been the pioneering work of {\bf Decio Levi} in this field \cite{levi1984,levi89,levi1989,levi1996,levi2005}. For the particular case of integrable PDEs, Lie analysis can be extended to the associated spectral problem, providing the symmetry transformations not only for the fields but also for the eigenfunctions and the spectral parameter \cite{albares2021}.
\end{itemize}

This article aims to combine the aforementioned approaches in order to analyze the complex version of the ubiquitous modified Korteveg-de Vries (mKdV) equation \cite{kirchev2010,Xu2023}, written as follows
\begin{equation}
\begin{aligned}
& u_t +u_{xxx}+6\overline uuu_x=0,\\
&\overline u_t + \overline u_{xxx}+6\overline uu\overline u_x=0,
\end{aligned}
\label{eq1}
\end{equation}
where $u=u(x,t)$ is a complex field and $\overline u(x,t)$ stands for its complex conjugate. Therefore, the modulus of this field can be written as
\begin{equation}
\left|u\right|^2=u\overline u.
\label{eq2}
\end{equation}
System (\ref{eq1}) can be alternatively written as 
\begin{eqnarray}\label{eq3}
&& u_t +u_{xxx}+6m_xu_x=0,\nonumber\\
&&\overline u_t + \overline u_{xxx}+6m_x\overline u_x=0,\\
&&m_x=u\overline u.
\nonumber
\end{eqnarray}
where we have introduced the real field $m(x,t)$ as the probability density.

This equation and its solutions have been studied in previous works from different perspectives \cite{Xu2023,Porsezian2014,Ma2022}. Our article aims to offer a unified study of this equation by means of the singular manifold method, so that the spectral problem can be algorithmically derived, as well as the construction of solutions.

Section \ref{sec:2} is devoted to the application of the SMM to the system of PDEs given in \eqref{eq3}, which allows us to compute its associated spectral problem, resulting in a two-component Lax pair. The SMM, when applied to the spectral problem itself, straightforwardly provides binary Darboux transformations. This procedure can be recursively iterated in order to derive an algorithmic description of the $\tau$-functions in terms of the eigenfunctions of the Lax pair. The construction of this approach can be found in Section \ref{sec:3}, together with its application to obtain interesting solutions of solitonic nature. Section \ref{sec:4} comprises the Lie symmetry analysis for the spectral problem. This method leads to the symmetry transformations for  the fields $u$, $\overline u$ as well as for the eigenfunctions and the spectral parameter. Nontrivial similarity reductions are exhaustively studied in Section \ref{sec:5}. These reductions enables us to determine spectral problems for ODEs, which are extraordinarily difficult to find through other methods. We conclude the paper with a section of conclusions.

\section{Singular manifold method}\label{sec:2}

The Painlev\'e expansion \cite{weiss} for the system (\ref{eq3}) means that the fields can be locally written as 
\begin{equation}
\begin{aligned}
u&=\sum_{j=0}^{\infty}a_j(x,t)\left[\phi(x,t)\right]^{j-\alpha},\\
\overline u&=\sum_{j=0}^{\infty}\overline a_j(x,t)\left[\phi(x,t)\right]^{j-\alpha},\\
m&=\sum_{j=0}^{\infty}b_j(x,t)\left[\phi(x,t)\right]^{j-\beta},
\end{aligned}
\label{eq4}
\end{equation}
where $\phi(x,t)=0$ is an arbitrary manifold, often denominated as the movable singularity manifold \cite{weiss}, and $\alpha,\beta\in\mathbb{N}$.

A leading-order balance easily yields
\begin{equation}
\alpha=1, \qquad\beta=1,\qquad a_0\overline a_0 =-\phi_x^2,\qquad  b_0=\phi_x.
\label{eq5}
\end{equation}

Substitution of the series (\ref{eq4}) into (\ref{eq3}) constitutes a cumbersome calculation, which can be easily computed with the aid of MAPLE. This process leads to recurrence relations among the coefficients $a_j,\,\overline{a}_j,\,b_j$, which can be expressed in matrix form as
\begin{equation}
\mathcal{A}\left( \begin{array}{c} 
a_j\\ 
\overline{a}_j\\
b_j
\end{array}
\right)=\mathcal{B},
\label{eq:5b}
\end{equation}
where $\mathcal{A}$ is the matrix
\begin{equation*}
\mathcal{A}=\left( \begin{array}{ccc} \phi_x^3j(j-1)(j-5) & 0 & -6\phi_x^2a_0(j-1)\\ 
0 & \phi_x^3j(j-1)(j-5)  &-6\phi_x^2\overline a_0(j-1)\\ -\overline a_0 & -a_0 & \phi_x(j-1)
\end{array}
\right),     
\end{equation*}
and the  vector $\mathcal{B}$ only depends on lower coefficients up to $a_{j-1},\,\overline{a}_{j-1},b_{j-1}$ and their derivatives. The values of $j$ which retrieve arbitrary coefficients in the series (\ref{eq4}) are called resonances. Given the resonance conditions \eqref{eq:5b}, it is easy to compute the resonances imposing the condition
\begin{equation*}
\operatorname{det}\mathcal{A}=\phi_x^7 (j+1) j(j-1)^2(j-3)(j-4)(j-5)=0.
\end{equation*}

Therefore the system exhibits simple resonances at $j=-1,0,3,4,5$, and a double resonance at $j=1$. The resonance $j=-1$ is associated with the arbitrariness of the singular manifold $\phi=0$. The remaining six resonances imply the presence of an arbitrary coefficient at each order of $j$ in the series expansions. The resonance conditions \eqref{eq:5b} are identically satisfied, allowing us to conclude that (\ref{eq3}) possesses the Painlevé Property, and it is therefore integrable \cite{wtc}. The resonance at $j=0$ allows us to introduce an arbitrary function $C(x,t)$ such that the leading coefficients in \eqref{eq5} can be written as
\begin{equation}
a_0=C\phi_x,\qquad \overline a_0=-\frac{\phi_x}{C}.
\label{eq6}
\end{equation}
\subsection{Truncated expansion}
The SMM \cite{weiss} focuses on truncated solutions of the Painlevé series (\ref{eq4}) in which all  the coefficients in positive powers of $\phi$ vanish, \textit{i.e.} solutions $\left\{u^{[1]},\,\overline u^{[1]},\,m^{[1]}\right\} $ of the form
\begin{equation}
\begin{aligned}
u^{[1]}&=a_1+C\frac{\phi_x}{\phi},\\
\overline u^{[1]}&=\overline a_1-\frac{1}{C}\frac{\phi_x}{\phi},\\
m^{[1]}&=b_1+\frac{\phi_x}{\phi}.
\end{aligned}
\label{eq7}
\end{equation}

Substitution of (\ref{eq7}) into (\ref{eq3}) gives rise to three polynomials in negative powers of $\phi$ whose coefficients should be zero, providing the following results:
\begin{itemize}

\item {\bf Auto-B\"acklund transformations}

The terms in $\phi^0$ indicate that the coefficients $\{a_1,\overline a_1,\, b_1\}$ are also solutions of the system (\ref{eq3}). Without loss of generality, we may therefore identify $\{a_1=u^{[0]},\, \overline a_1=\overline u^{[0]},\, b_1=m^{[0]}\}$ such that expression (\ref{eq7}) corresponds to the auto-B\"acklund transformation
\begin{equation}\label{eq8}
\begin{aligned}
u^{[1]}&=u^{[0]}+C\frac{\phi_x}{\phi},\\
\overline u^{[1]}&=\overline u^{[0]}-\frac{1}{C}\frac{\phi_x}{\phi},\\
m^{[1]}&=m^{[0]}+\frac{\phi_x}{\phi},
\end{aligned}
\end{equation}
 between two solutions $\left\{u^{[0]},\overline u^{[0]}, m^{[0]}\right\}$ and $\{u^{[1]}, \overline u^{[1]},m^{[1]}\}$ of system (\ref{eq3}), where the first solution is called the seed solutions and the second solution is the iterated one. Obviously, $m^{[0]}$ and $m^{[1]}$ can be also expressed as
\begin{equation*}
m^{[0]}_x=u^{[0]}\overline u^{[0]},\qquad m^{[1]}_x=u^{[1]}\overline u^{[1]}.
\end{equation*}

\item{\bf Seed solutions}

The lower-order powers in $\phi$ of the three equations in (\ref{eq3}) are 

\begin{eqnarray}
&& -\frac{C\phi_x^2}{\phi^2}\left[6\frac{u^{[0]}_x}{C}+6m_x+R+4\frac{\phi_{xxx}}{\phi_x}-3\left(\frac{\phi_{xx}}{\phi_x}\right)^2+3\frac{C_{xx}}{C}+3\frac{\phi_{xx}}{\phi_x}\frac{C_x}{C}\right]=0,\nonumber\\
&&\,\,\,\,\,\frac{\phi_x^2}{C\phi^2}\left[-6C\,\overline u^{[0]}_x+6m_x+R+4\frac{\phi_{xxx}}{\phi_x}-3\left(\frac{\phi_{xx}}{\phi_x}\right)^2-3\frac{C_{xx}}{C}+6\frac{C_x^2}{C^2}-3\frac{\phi_{xx}}{\phi_x}\frac{C_x}{C}\right]=0,\nonumber\\
&&\quad\,\,\,\frac{\phi_x}{\phi}\left[\frac{u^{[0]}}{C}-C\,\overline u^{[0]}+\frac{\phi_{xx}}{\phi_x}\right]=0,
\nonumber
\end{eqnarray}
whose combination permits to express the seed fields in terms of the functions $\phi$ and $C$ as
\begin{equation}
\begin{aligned}
u^{[0]}&=\frac{C}{2}\left(-\frac{C_x}{C}-\frac{\phi_{xx}}{\phi_x}-\lambda\right),\\
\overline u^{[0]}&=\frac{1}{2C}\left(-\frac{C_x}{C}+\frac{\phi_{xx}}{\phi_x}-\lambda\right),
\end{aligned}
\label{eq9}
\end{equation}
where $\lambda$ appears as an integration constant.

\item {\bf Singular manifold equations}

The remaining powers in $\phi$ establish relations between the functions $C$ and $\phi$, known as the singular manifold equations, which read
\begin{equation}
\begin{aligned}
R&=-V_x+\frac{V^2}{2}-\frac{3}{2}\left(\lambda^2+\frac{C_x^2}{C^2}\right),\\
\frac{C_t}{C}&=-\frac{C_{xxx}}{C}+\frac{3\,C_{xx}C_x}{C^2}-\frac{9\,C_{x}^3}{2C^3}-3\left(V_x-\frac{V^2}{2}+\frac{\lambda^2}{2}\right)\frac{C_x}{C}-3\lambda\frac{C_x^2}{C^2},\\
V_t&=\left(R_x+VR\right)_x,
\end{aligned}
\label{eq10}
\end{equation}
where $V,R$ are defined as
\begin{equation}
 V=\frac{\phi_{xx}}{\phi_x},\qquad
R=\frac{\phi_{t}}{\phi_x}.
\label{eq11}
\end{equation}
\end{itemize}

\subsection{Spectral problem}

Equations (\ref{eq9}) for the fields can be straightforwardly linearized by redefining $\phi$ and $C$ in terms of two auxiliary functions $\psi$ and $\chi$ in the form
\begin{equation}
\phi_x=\psi\chi,\qquad C=\displaystyle{\frac{\psi}{\chi}},
\label{eq11a}
\end{equation}
providing the following relations
\begin{equation}
V=\displaystyle{\frac{\psi_x}{\psi}+\frac{\chi_x}{\chi}},\qquad
R=-\displaystyle{\frac{3}{2}\lambda^2-\frac{\psi_{xx}}{\psi}-\frac{\chi_{xx}}{\chi}+4\frac{\psi_x}{\psi}\frac{\chi_x}{\chi}}.
\label{eq12}    
\end{equation}

Thus, expressions (\ref{eq9}) read  
\begin{equation}
\begin{aligned}
&\psi_x=-\frac{\lambda}{2}\psi-u^{[0]}\chi,\\
&\chi_x=\frac{
\lambda}{2}\chi+\overline u^{[0]}\psi,
\end{aligned}
\label{eq13}
\end{equation}
whilst equations from (\ref{eq10}) take the form
\begin{equation}
\begin{aligned}
&\frac{\psi_t}{\psi}-\frac{\chi_t}{\chi}=-\frac{C_{xxx}}{C}+\frac{3\,C_{xx}C_x}{C^2}-\frac{9\,C_{x}^3}{2C^3}-3\left(V_x-\frac{V^2}{2}+\frac{\lambda^2}{2}\right)\frac{C_x}{C}-3\lambda\frac{C_x^2}{C^2},\\
&\frac{\psi_t}{\psi}+\frac{\chi_t}{\chi}=R_x+VR.
\end{aligned}
\label{eq14}
\end{equation}

After introducing (\ref{eq11a}) and (\ref{eq12}), equations \eqref{eq14} result in
\begin{equation}
\begin{aligned}
&\psi_t=\left(\,\frac{\lambda^3}{2}+\lambda u^{[0]}\overline u^{[0]}+u^{[0]}\overline u^{[0]}_x-\overline u^{[0]} u^{[0]}_x\right)\psi\,+\left(+\lambda^2 u^{[0]}-\lambda u^{[0]}_x+2 \overline u^{[0]}\left (u^{[0]}\right)^2 +u^{[0]}_{xx}\right)\chi,\\
&\chi_t=\left(-\frac{\lambda^3}{2}-\lambda u^{[0]}\overline u^{[0]}+\overline u^{[0]} u^{[0]}_x-u^{[0]}\overline u^{[0]}_x\right)\chi+\left(-\lambda^2 \overline u^{[0]}-\lambda \overline u^{[0]}_x-2u^{[0]}\left( \overline u^{[0]}\right)^2-
\overline u^{[0]}_{xx}\right)\psi.
\end{aligned}
\label{eq15}
\end{equation}

Thus, equations (\ref{eq13}) and (\ref{eq15}) constitute the spectral problem of the complex mKdV equation, which can be written in matrix form as
\begin{equation}
\begin{aligned}
\vec\Psi_x&=-\left(\frac{\lambda }{2}A+B_{\left\{u^{[0]},\overline u^{[0]}\right\}}\right)\vec\Psi,\\
\vec\Psi_t&=\left(\frac{\lambda^3}{2}A+\lambda^2 B_{\left\{u^{[0]},\overline u^{[0]}\right\}}+\lambda\left[\left(-D_{\left\{u^{[0]},\overline u^{[0]}\right\}}\right)_x+u^{[0]}\overline u^{[0]}A\right]\right)\vec\Psi\\
&+\left(\left[B_{\left\{u^{[0]},\overline u^{[0]}\right\}}\right]_{xx}+2u^{[0]}\overline u^{[0]}B_{\left\{u^{[0]},\overline u^{[0]}\right\}}+\left[\overline u^{[0]} u^{[0]}_x-u^{[0]}\overline u^{[0]}_x\right]A\right)\vec\Psi.
 \end{aligned}\label{eq16}\end{equation}
where $\vec\Psi=(\psi,\chi)^\intercal$ and $A, B, D$ are the matrices
\begin{equation}
A=\left( \begin{array}{cc} 1&0
 \\ 0 & -1
		\end{array}
		\right), \quad B_{\left\{u,\overline u\right\}}= \left( \begin{array}{cc} 0&u
 \\ -\overline u & 0 
		\end{array}
		\right), \quad D_{\left\{u,\overline u\right\}}= \left( \begin{array}{cc} 0&u
 \\ \overline u & 0 
		\end{array}
		\right).\label{eq17}
 \end{equation}
The relations among the function $C$, the singular manifold $\phi$ and the eigenfunctions $\psi,\chi$ follow from (\ref{eq11})-(\ref{eq12}), yielding
 \begin{equation}
 \begin{aligned}
C&=\frac{\psi}{\chi},\\
d\phi&=\displaystyle{\psi\chi dx+\left(-\frac{3}{2}\lambda^2\psi\chi-\psi_{xx}\chi-\chi_{xx}\psi+4\psi_x\chi_x\right)dt},\label{eq18}
 \end{aligned}\end{equation}
while the auto-Bäcklund transformation (\ref{eq8}) takes the form 
\begin{equation}
\begin{aligned}
 u^{[1]}&=u^{[0]}+\frac{\psi^2}{\phi},\\
\overline u^{[1]}&=\overline u^{[0]}-\frac{\chi^2}{\phi},\\
m^{[1]} &= m^{[0]}+\frac{\phi_x}{\phi}.
\end{aligned}
\label{eq19}
\end{equation}

\section{Darboux transformations} \label{sec:3}

Truncated expansion (\ref{eq19}) can be understood as an auto-B\"acklund transformation between the seed solution $\{u^{[0]},\overline u^{[0]},m^{[0]}\}$ and the iterated solution $\{u^{[1]},\overline u^{[1]},m^{[1]}\}$. The Lax pair for this last solution, of spectral parameter $\lambda_2$, has the expression 
 \begin{equation}
 \begin{aligned}
(\vec\Psi_{1,2})_x&=-\left(\frac{\lambda_2 }{2}A+B_{\left\{u^{[1]},\overline u^{[1]}\right\}}\right)\vec\Psi_{1,2},\\(\vec\Psi_{1,2})_t&=\left(\frac{\lambda_2^3}{2}A+\lambda_2^2 B_{\left\{u^{[1]},\overline u^{[1]}\right\}}+\lambda_2\left[\left(-D_{\left\{u^{[1]},\overline u^{[1]}\right\}}\right)_x+u^{[1]}\overline u^{[1]}A\right]\right)\vec\Psi_{1,2}\\
&+\left(\left[B_{\left\{u^{[1]},\overline u^{[1]}\right\}}\right]_{xx}+2u^{[1]}\overline u^{[1]}B_{\left\{u^{[1]},\overline u^{[1]}\right\}}+\left[\overline u^{[1]} u^{[1]}_x-u^{[1]}\overline u^{[1]}_x\right]A\right)\vec\Psi_{1,2},
 \end{aligned}\label{eq20}\end{equation}
where  $\vec\Psi_{1,2} = (\psi_{1,2},\chi_{1,2})^\intercal$ is the eigenvector associated with the iterated solution. 
 
The Lax pair is frequently treated as a linear system for the eigenfunctions. Nevertheless, another approach to tackle this problem is to consider (\ref{eq20}) as a set of nonlinear coupled differential equations among the fields and eigenfunctions \cite{konopelchenko}. Consequently, the truncation of the Painlevé series for the fields must be accompanied by a similar truncation for the eigenfunctions, of the form
  \begin{equation}
 \begin{aligned}
  u^{[1]}&=u^{[0]}+\frac{\psi_1^2}{\phi_1},\\
\overline u^{[1]}&=\overline u^{[0]}-\frac{\chi_1^2}{\phi_1},\\
m^{[1]}&=m^{[0]}+\frac{(\phi_1)_x}{\phi_1},\\
\psi_{1,2}&=\psi_2-\psi_1\frac{\Delta_{1,2}}{\phi_1},\\
\chi_{1,2}&=\chi_2-\chi_1\frac{\Delta_{1,2}}{\phi_1},
 \end{aligned}\label{eq21}\end{equation}
where $\vec\Psi_i=(\psi_i,\chi_i)^\intercal,\,i=1,2$, is an eigenvector for the seed Lax pair (\ref{eq16}) of eigenvalue $\lambda_i$, and $\Delta_{1,2}=\Delta_{1,2}(x,t)$ is a complex function to be determined. Each eigenvector $\vec\Psi_i$ then defines a singular manifold $\phi_i$ through (\ref{eq18}) as
 \begin{equation}
 d\phi_i=\displaystyle{\psi_i\chi_i dx+\left(-\frac{3}{2}\lambda_i^2\psi_i\chi_i-(\psi_i)_{xx}\chi_i-(\chi_i)_{xx}\psi_i+4(\psi_i)_x(\chi_i)_x\right)dt}.\label{eq22}
\end{equation}
Substitution of (\ref{eq21}) in (\ref{eq20}) provides the following expression for $\Delta_{1,2}$ in terms of the eigenvectors $\vec\Psi_i$, 
\begin{equation}
\Delta_{1,2}=\frac{\psi_1\chi_2-\psi_2\chi_1}{\lambda_2-\lambda_1}.\label{eq23}
\end{equation}

An analogous reasoning allows us to regard (\ref{eq22}) as a nonlinear equation that links eigenfunctions and singular manifolds, such that the truncation for the eigenfunctions leads to the following truncated expansion for the singular manifold
\begin{equation}
\phi_{1,2}=\phi_2-\frac{\Delta_{1,2}^2}{\phi_1}.\label{eq24}
\end{equation}
The truncated Painlevé expansion series (\ref{eq21}) can therefore be considered as binary Darboux transformations \cite{matveev} such that both fields and eigenfunctions for the iterated Lax pair (\ref{eq20}) are constructed in terms of two solutions for the seed Lax pair (\ref{eq16}), associated with two different values of the spectral parameter,  $\lambda_1$ and $\lambda_2$, respectively.

\subsection{Iteration of solutions}

The iterated singular manifold (\ref{eq24}) can be employed to generate a new iteration of the fields in the form
\begin{equation}
\begin{aligned}
u^{[2]}&=u^{[1]}+\frac{\psi_{12}^2}{\phi_{1,2}},\\
\overline u^{[2]}&=\overline u^{[1]}-\frac{\chi_{1,2}^2}{\phi_{1,2}},\\
\overline m^{[2]}&=m^{[1]}+\frac{(\phi_{1,2})_x}{\phi_{1,2}},\\
\end{aligned}\label{eq25}
\end{equation}
which in combination with (\ref{eq21}) leads to
\begin{equation}
\begin{aligned}
u^{[2]}&=u^{[0]}+\frac{\psi_2^2\phi_1+\psi_1^2\phi_2-2\psi_1\psi_2\Delta_{1,2}}{\tau_{1,2}},\\
\overline u^{[2]}&=\overline u^{[0]}-\frac{\chi_2^2\phi_1+\chi_1^2\phi_2-2\chi_1\chi_2\Delta_{1,2}}{\tau_{1,2}},\\
m^{[2]}&=m^{[0]}+\frac{(\tau_{1,2})_x}{\tau_{1,2}},
\end{aligned}\label{eq26}
\end{equation}
where now the $\tau$-function is defined as
\begin{equation}
\tau_{1,2}=\phi_1\phi_{1,2}=\phi_1\phi_2-\Delta_{1,2}^2. 
\label{eq27}
\end{equation}
Generally, the $n$th iteration for the fields can be easily constructed as
\begin{equation}
\begin{aligned}
m^{[n]}&= m^{[0]}+\frac{(\tau_{1,2,\dots,n})_x}{\tau_{1,2,\dots,n}},\\
\tau_{1,2,\dots,n}&=\det \Delta_{i,j},
\end{aligned}\label{eq28}
\end{equation}
where $\tau_{1,2,\dots,n}$ constitutes the generalization of the $\tau$-function in \eqref{eq27} and $\Delta _{i,j}$ is given by the matrix
\begin{equation}
\begin{aligned}
\Delta _{i,i}&=\phi_i,\\
\Delta _{i,j}&=\frac{\psi_i\chi_j-\psi_j\chi_i}{\lambda_j-\lambda_i} ,\quad i,j=1,\dots,n.
\end{aligned}\label{eq29}
\end{equation}

It is worthwhile to remark that $\tau_{1,2,\dots,n}$ is the same $\tau$-function that appears in Hirota's direct method \cite{Hirota}. Thus, the SMM allows us not only to identify the spectral problem (\ref{eq16}), but also to determine many key properties related to integrability, such as auto-B\"acklund transformations (\ref{eq19}), Darboux transformations (\ref{eq21}) and $\tau$-functions (\ref{eq29}). Moreover, it also provides an algorithmic procedure to obtain solutions in a more compact and simple way than other direct methods \cite{kirchev2010,Xu2023,Porsezian2014,Ma2022}.

\subsection{Soliton solutions}

We start with a trivial plane wave as seed solution, of the form                      
\begin{equation}
u^{[0]}=j_0E_0,\qquad \overline u^{[0]}=\frac{j_0}{E_0},\qquad
E_0=E_0(x,t)=e^{ik_0\left[(x+(k_0^2-6j_0^2)t\right]},
\label{eq30}
\end{equation}
where $k_0$ and $j_0$ are arbitrary constants.

The seed  eigenfunctions for the Lax pair (\ref{eq16}) then read              
\begin{equation}\psi_j=e^{\theta_j}\frac{E_jE_0^{\frac{1}{2}}}{H_j},   \qquad  \chi_j=e^{-\theta_j}\frac{E_j}{E_0^{\frac{1}{2}}H_j},\qquad j=1,2,\dots,n,
\label{eq31}
\end{equation}
where
\begin{equation*}
E_j=E_j(x,t)=e^{k_j\left[x+v_jt\right]},\qquad H_j=H_j(t)=e^{ic_jt}.
\end{equation*}
The spectral parameter $\lambda_j$ and the constants $k_j,v_j,c_j$ can be expressed in terms of the sole arbitrary parameter $\theta_j$ through the following identities
\begin{equation}
\begin{aligned}
\lambda_j&=-ik_0-2j_0\cosh(2\theta_j),\\
k_j&=j_0\sinh(2\theta_j),\\
v_j&=3k_0^2-6j_0^2-4j_0^2\sinh^2(2\theta_j),\\
c_j&=3k_jk_0j_0\cosh(2\theta_j).
\end{aligned}\label{eq32}
\end{equation}

The singular manifolds and the $\Delta$-matrix are easily obtained from (\ref{eq22}) and (\ref{eq29}), yielding
\begin{equation}
\phi_i=\frac{1}{2k_i}\left(d_i+\frac{E_i^2}{H_i^2}\right),\qquad
\Delta_{i,j}=\frac{1}{2j_0\sinh(\theta_i+\theta_j)}\frac{E_iE_j}{H_iH_j},\qquad i,j=1,2,\dots,n,
\label{eq33}
\end{equation}
where $d_i$ are arbitrary constants.

These results forthrightly lead to the $n$-soliton solution for \eqref{eq3} through (\ref{eq28}). In the case of two solitons, the $\tau$-function explicitly reads 
\begin{equation}
\tau_{1,2}=\frac{1}{4k_1k_2}\left[d_1d_2+\frac{d_1E_2^2}{H_2^2}+\frac{d_2E_1^2}{H_1^2}+A_{1,2}\frac{E_1^2E_2^2}{H_1^2H_2^2}\right],\\
\label{eq34}
\end{equation}
where
\begin{equation}
A_{1,2}=1-\frac{\sinh(2\theta_1)\sinh(2\theta_2)}{\sinh^2(\theta_1+\theta_2)}.\label{eq35}
\end{equation}

Taking into account (\ref{eq2}) and (\ref{eq28}), the two-soliton solution of (\ref{eq1}) is written as
\begin{equation}
\left|u^{[2]}\right|^2=m^{[2]}_x=j_0^2+\left(\frac{(\tau_{1,2})_x}{\tau_{1,2}}\right)_x,
\end{equation}
with $\tau_{1,2}$ given in (\ref{eq34}). 

In the following, we  explore various types of solutions depending on different values of the free parameters $k_0$, $\theta_1$, $\theta_2$.

\subsection{Line solitons}

The usual soliton solutions can be obtained by imposing $k_0=0$. This choice necessarily implies that $H_1=H_2=1$. Then, considering the center-of-mass reference frame, which follows from the Galilean transformation
\begin{equation*}
x=X-\frac{v_1+v_2}{2}t,
\end{equation*}
the $\tau$-function in \eqref{eq34} becomes
\begin{equation}
\tau_{1,2}=\frac{1}{4k_1k_2}\left[d_1d_2+d_1e^{2k_2(X-Vt)}+d_2e^{2k_1(X+Vt)}+A_{1,2}e^{2k_2(X-Vt)}e^{2k_1(X+Vt)}\right],\\
\label{eq37}
\end{equation}
where $d_1,\,d_2$ are arbitrary parameters, and $V$ is the relative velocity of the two solitons given by
\begin{equation*}
V=\frac{v_1-v_2}{2}=2j_0^2(\sinh^2(2\theta_2)-\sinh^2(2\theta_1)).
\end{equation*}
Similar solutions have been derived in \cite{Xu2023} through Hirota's bilinear method \cite{Hirota}.

\begin{figure}[H]
\centering
\includegraphics[width=0.45\columnwidth]{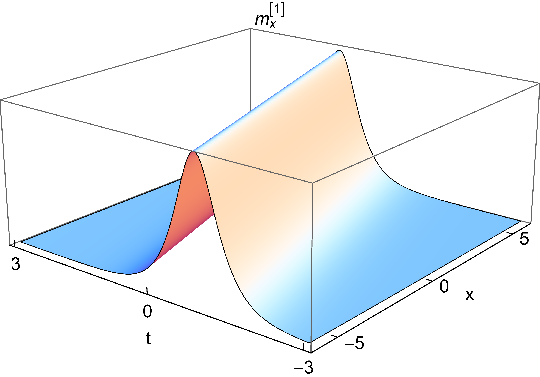} 
\hspace{0.08cm}
\includegraphics[width=0.45\columnwidth]{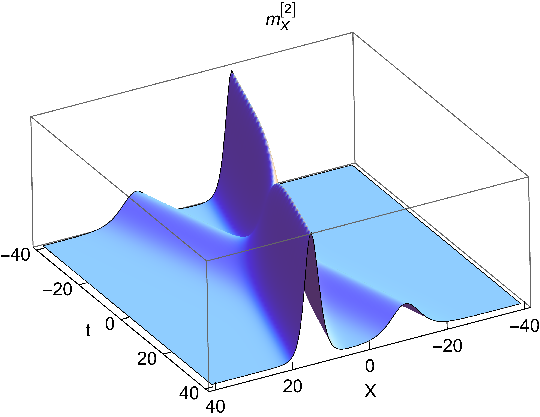}
\caption{One-soliton solution (left) and two-soliton solution (right) for $k_0=0,j_0=1,d_1=1,d_2=2,\theta_1=0.1,\theta_2=0.2.$}
\label{fig:1}
\end{figure}

 Figure \ref{fig:1}  displays the one-soliton solution (left) and the two-soliton solution (right) for the system \eqref{eq3}. It can be straightforwardly seen that the first iteration for the fields provides the usual line soliton (left), whilst the second iteration allows us to construct the usual scattering between two solitons (right), where the sole effect of their interactions is a phase shift in the direction of propagation of the solitary waves.

\subsection{Breathers}
\subsubsection*{Temporal breathers}
Choosing $\theta_2=\frac{i\pi}{2}-\theta_1$ and $d_1=d_2=\cosh(2\theta_1)$, the different parameters defined in \eqref{eq32} read
\begin{equation}
\begin{aligned}
\lambda_1&=-\overline{\lambda_2}=-ik_0-2j_0\cosh(2\theta_1),\\
k_1&=k_2=j_0\sinh(2\theta_1),\\
v_1&=v_2=3k_0^2-6j_0^2-4j_0^2\sinh^2(2\theta_1),\\
c_1&=-c_2=3k_0j_0^2\sinh(4\theta_1),
\end{aligned}
\label{eq38}
\end{equation}
such that the $\tau$-function from \eqref{eq34} can be expressed as
\begin{equation}
\tau_{1,2}=\frac{\cosh(2\theta_1)E_1^2}{2k_1^2}\left[\cosh\left(2k_1(x+v_1t))\cosh(2\theta_1\right)+\cos(2c_1t)\right],
\label{eq39}
\end{equation}
which is a solution with a periodic behaviour in time but hiperbolic in space. Solution \eqref{eq39} can be considered as the equivalent of the Kuznetsov-Ma breather from NLS equation \cite{kuznetsov,ma1979}.

\subsubsection*{Spatial breathers}
Taking $\theta_1=i\varphi_1$ as a purely imaginary parameter in (\ref{eq38}), it is easy to get
\begin{equation}
\begin{aligned}
\lambda_1&=-ik_0-2j_0\cos(2\varphi_1),\\
k_1&=iK_1=ij_0\sin(2\varphi_1),\\
v_1&=3k_0^2-6j_0^2+4j_0^2\sin^2(2\varphi_1),\\
c_1&=iC_1=3ik_0j_0^2\sin(4\varphi_1),
\end{aligned}
\label{eq40}
\end{equation}
such that the $\tau$-function now reads
\begin{equation}
\tau_{1,2}=-\frac{\cos(2\varphi_1)E_1^2}{2K_1^2}\left[\cos\left(2K_1(x+v_1t))\cos(2\varphi_1\right)+\cosh(2C_1t)\right].
\end{equation}
This $\tau$-function is oscillatory in space and hyperbolic in time, yielding a solution  analogous to the Akhmediev breather from NLS equation \cite{akhmediev86}.

\begin{figure}[H]
\centering
\includegraphics[width=0.45\columnwidth]{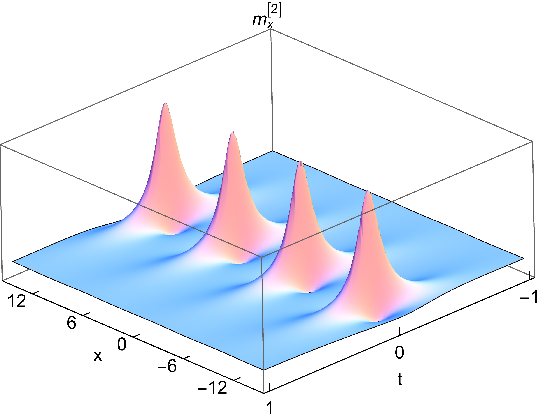} 
\hspace{0.08cm}
\includegraphics[width=0.45\columnwidth]{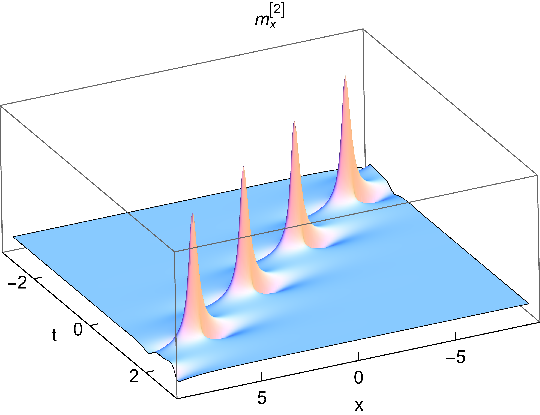}
\caption{Analogous Akhmediev breather (left) and Kuznetsov-Ma breather (right) for the complex mKdV equation, with parameter $k_0=1,j_0=1,\theta_1=0.2,\varphi_1=0.2$.}
\label{fig:2}
\end{figure}

The behaviour of these breather solutions has been plotted in Figure \ref{fig:2}.

\subsection{Rogue waves}

Rogue waves are known as rational soliton solutions that are localized in space and time \cite{akhmediev2009}. This kind of solutions can be easily obtained taking the ansatz $\theta_1=0, \theta_2=i\frac{\pi}{2}$, such that the eigenfunctions for the seed Lax pair exhibit now polynomial expressions of the form
\begin{equation}
\begin{aligned}
\psi_1&=E_0^{\frac{1}{2}}\left[a\left(X-6ik_0j_0t+\frac{1}{2j_0}\right)+z_0\right],   \quad  &\chi_1&=\frac{1}{E_0^{\frac{1}{2}}}\left[a\left(X-6ik_0j_0t-\frac{1}{2j_0}\right)+z_0\right],\\
\psi_2&=iE_0^{\frac{1}{2}}\left[a\left(X+6ik_0j_0t-\frac{1}{2j_0}\right)+z_0\right],   \quad  &\chi_2&=-\frac{i}{E_0^{\frac{1}{2}}}\left[a\left(X+6ik_0j_0t+\frac{1}{2j_0}\right)+z_0\right],\label{eq42}
\end{aligned}
\end{equation}
in the comoving rest frame $X=x+3(k_0^2-2j_0^2)t$. Parameters $a,\,z_0$ are arbitrary constants.
 
The spectral parameters $\lambda_1$ and $\lambda_2$ are no longer arbitrary but rather related to the parameters of the seed solution in the form
\begin{equation}
\lambda_1=-ik_0-2j_0,\qquad \lambda_2=-ik_0+2j_0.\label{eq43}
\end{equation}
The singular manifold and $\Delta$-matrix follow from (\ref{eq22}) and (\ref{eq23})
\begin{equation}
\begin{aligned}
\phi_1=&\left[\frac{a^2X^3}{3}+(aX+z_0)\left(z_0X-36ak_0^2j_0^2t^2\right)-\frac{a^2}{4j_0^2}\left(X-16j_0^2t\right)+a_0\right]\\
+i&\left[6k_0j_0t\left(12a^2k_0^2j_0^2t^2-(aX+z_0)^2+\frac{3a^2}{4j_0^2}\right)+b_0\right],\\
\phi_2=&\left[\frac{a^2X^3}{3}+(aX+z_0)\left(z_0X-36ak_0^2j_0^2t^2\right)-\frac{a^2}{4j_0^2}\left(X-16j_0^2t\right)+a_0\right]\\
-i&\left[6k_0j_0t\left(12a^2k_0^2j_0^2t^2-(aX+z_0)^2+\frac{3a^2}{4j_0^2}\right)+b_0\right],\\
\Delta_{1,2}=-\frac{i}{2j_0}&\left[\left(aX+z_0\right)^2+36a^2k_0^2j_0^2t^2+\frac{a^2}{4j_0^2}\right].
\end{aligned}
\label{eq44}
\end{equation}
It is worthwhile to notice that $\phi_2$ is the complex conjugate of $\phi_1$.

Then, the second iteration for the fields provides
\begin{equation}
\left|u^{[2]}\right|^2=j_0^2+\left(\frac{(\tau_{1,2})_x}{\tau_{1,2}}\right)_x,\label{eq45}
\end{equation}
where $\tau_{1,2}$ is now given by 
\begin{equation}
\begin{aligned}
\tau_{1,2}=&\left[\frac{a^2X^3}{3}+(aX+z_0)\left(z_0X-36ak_0^2j_0^2t^2\right)-\frac{a^2}{4j_0^2}\left(X-16j_0^2t\right)+a_0\right]^2+\\+&\left[6k_0j_0t\left(12a^2k_0^2j_0^2t^2-(aX+z_0)^2+\frac{3a^2}{4j_0^2}\right)+b_0\right]^2\\+\frac{1}{4j_0^2}&\left[\left(aX+z_0\right)^2+36a^2k_0^2j_0^2t^2+\frac{a^2}{4j_0^2}\right]^2.
\end{aligned}
\label{eq46}
\end{equation}

This solution, expressed in terms of six arbitrary constants $a,a_0,b_0,z_0,j_0,k_0$, generalizes other solutions previously obtained in \cite{Xu2023}. Since $\tau$-function (\ref{eq46}) has no zeroes, then solution (\ref{eq45}) is free of singularities. This solution includes two different cases:

\begin{enumerate}

\item $a=0,\, z_0=1$.

In this case, the $\tau$-functions \eqref{eq46} reads
\begin{equation}
\begin{aligned}
\tau_{1,2}=&\left[X+a_0\right]^2+\left[-6k_0j_0t+b_0\right]^2+\frac{1}{4j_0^2},
\end{aligned}
\label{eq47}
\end{equation}
which provides the equivalent solution of the Peregrine soliton for the NLS equation \cite{peregrine}.

\item $a=1,\, z_0=0$.

This choice for the parameters yields a second family of rogue waves for system \eqref{eq1} with a more intricate behaviour. The $\tau$-function therefore reads
\begin{equation}
\begin{aligned}
\tau_{1,2}&=\left[\left(\frac{X^3}{3}-36k_0^2j_0^2t^2X-\frac{X}{4j_0^2}+4t\right)+a_0\right]^2\\
&+\left[-6k_0j_0t\left(X^2-12k_0^2j_0^2t^2-\frac{3}{4j_0^2}\right)+b_0\right]^2+\frac{1}{4j_0^2}\left[X^2+36j_0^2k_0^2t^2+\frac{1}{4j_0^2}\right]^2,
\end{aligned}
\label{eq48}
\end{equation}
which provides a solution that asymptotically behaves as two rogue waves moving along the curves
\begin{equation*}
\left(\frac{X^3}{3}-36k_0^2j_0^2t^2X-\frac{X}{4j_0^2}+4t\right)+a_0=0,
\end{equation*}
and
\begin{equation*}
-6k_0j_0t\left(X^2-12k_0^2j_0^2t^2-\frac{3}{4j_0^2}\right)+b_0=0,
\end{equation*}
respectively.

\end{enumerate}

\begin{figure}[H]
\centering
\includegraphics[width=0.45\columnwidth]{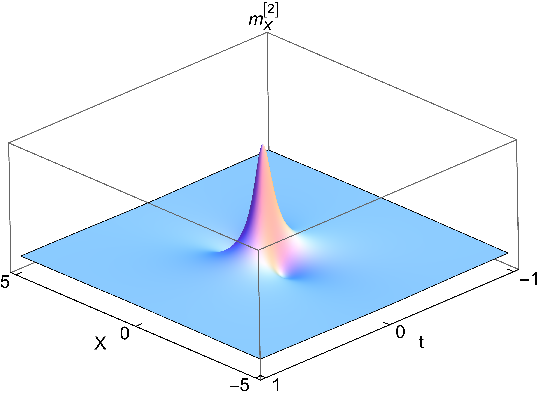} 
\hspace{0.08cm}
\includegraphics[width=0.45\columnwidth]{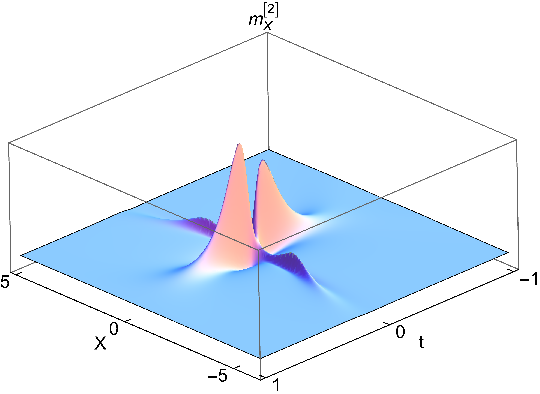}
\caption{Rogue waves for the complex mKdV equation when $a=0,z_0=1$ (left) and $a\neq 0,z_0=0$ (right), for parameters $k_0=1,j_0=1,a_0=0,b_0=0$.}
\label{fig:3}
\end{figure}

Figure \ref{fig:3} depicts the spatio-temporal behaviour of these solutions for both cases, when $a=0$ and $a\neq 0$.

\section{Lie symmetries}\label{sec:4}

Group analysis constitutes another fundamental technique to analyze any sort of problems involving differential equations \cite{ovsiannikov1992,lie1888}. The Lie symmetry method and the construction of group invariant solutions provide a consolidated framework to approach integrable systems. It  allows us to analyze not only the nonlinear PDEs of interest but also their associated spectral problems, yielding valuable information about the integrability structure of such systems. The novelty in our approach lies in considering simultaneous group transformations for the spectral parameter as a new independent variable for the eigenfunctions of the spectral problem. This  retrieves information about how both eigenfunctions and the spectral parameter are transformed under the action of the symmetry group \cite{albares2019,albares2021,estevez2005,estevez2013}, together with the usual symmetry transformations for the independent variables and the fields. Moreover, as the Lax pair constitutes an equivalent system to the former nonlinear equation, the symmetries of the spectral problem are expected to include the symmetries of the initial system \cite{legare1996}. 
This Section therefore concerns the application of Lie's method \cite{olver1986,stephani1990} to the spectral problem, in order to obtain classical symmetries for such system of PDEs. 

Let us focus on the spectral problem (\ref{eq16})
 \begin{equation}
 \begin{aligned}
\vec\Psi_x&=-\left(\frac{\lambda }{2}A+B_{\left\{u,\overline u\right\}}\right)\vec\Psi,\\\vec\Psi_t&=\left(\frac{\lambda^3}{2}A+\lambda^2 B_{\left\{u,\overline u\right\}}+\lambda\left[\left(-D_{\left\{u,\overline u\right\}}\right)_x+u\overline uA\right]\right)\vec\Psi\\&+\left(\left[B_{\left\{u,\overline u\right\}}\right]_{xx}+2u\overline uB_{\left\{u,\overline u\right\}}+\left[\overline u u_x-u\overline u_x\right]A\right)\vec\Psi,
 \end{aligned}\label{eq:1}\end{equation}
related to the system
  \begin{equation}
 \begin{aligned}
& u_t +u_{xxx}+6u\overline uu_x=0,\\
&\overline u_t + \overline u_{xxx}+6u\overline u \overline u_x=0.
\end{aligned}\label{eq:2}
\end{equation}
Let us consider the following $\epsilon$-parametric group of infinitesimal transformations
\begin{equation}
\begin{aligned}
\tilde{x}&=x+\epsilon\,\xi_x(x,t,\lambda,u,\overline{u},\psi,\chi)+\mathcal{O}(\epsilon^2),\\
\tilde{t}&=t+\epsilon\,\xi_t(x,t,\lambda,u,\overline{u},\psi,\chi)+\mathcal{O}(\epsilon^2),\\
\tilde{\lambda}&=\lambda+\epsilon\,\xi_\lambda(x,t,\lambda,u,\overline{u},\psi,\chi)+\mathcal{O}(\epsilon^2),\\
\tilde{u}&=u+\epsilon\,\eta_{u}(x,t,\lambda,u,\overline{u},\psi,\chi)+\mathcal{O}(\epsilon^2),\\
\tilde{\overline{u}}&=\overline{u}+\epsilon\,\eta_{\overline{u}}(x,t,\lambda,u,\overline{u},\psi,\chi)+\mathcal{O}(\epsilon^2),\\
\tilde{\psi}&=\psi+\epsilon\,\eta_\psi(x,t,\lambda,u,\overline{u},\psi,\chi)+\mathcal{O}(\epsilon^2),\\
\tilde{\chi}&=\chi+\epsilon\,\eta_\chi(x,t,\lambda,u,\overline{u},\psi,\chi)+\mathcal{O}(\epsilon^2),
\end{aligned}
\label{eq:3}
\end{equation}
generated by the vector field
\begin{equation}
X=\xi_x\partial_x+\xi_t\partial_t+\xi_{\lambda}\partial_{\lambda}+\eta_{u}\partial_{u}+\eta_{\overline u}\partial_{\overline u}+\eta_{\psi}\partial_{\psi}+\eta_{\chi}\partial_{\chi},
\label{eq:4}
\end{equation}
where $\epsilon$ is the group parameter and $\xi_x,\,\xi_t,\,\xi_{\lambda},\,\eta_{u},\,\eta_{\overline{u}},\,\eta_{\psi},\,\eta_{\chi}$ are the infinitesimals associated with the independent variables $x,t$, the spectral parameter $\lambda$, the fields $u,\overline u$ and the eigenfunctions $\psi,\chi$, respectively. Transformation \eqref{eq:3} can be easily extended to the different derivatives of the dependent variables \cite{olver1986,stephani1990}, such that the spectral problem \eqref{eq:1} remains invariant, resulting in the following set of symmetries
\begin{equation}
\begin{aligned}
\xi_x&=k_1x+k_2,\\
\xi_t&=3k_1t+k_3,\\
\xi_\lambda&=-k_1\lambda,\\
\eta_{u}&=\left(-k_1+2ik_4\right)u,\\
\eta_{\overline{u}}&=\left(-k_1-2ik_4\right)\overline u,\\
\eta_{\psi}&=(K(\lambda)+ik_4)\,\psi,\\
\eta_{\chi}&=(K(\lambda)-ik_4)\,\chi,\\
\end{aligned}
\label{eq:5}
\end{equation}
where $k_i,\,i=1,\dots,4$ are arbitrary real constants and $K(\lambda)$ is an arbitrary real function of $\lambda$. 

\subsection{Lie algebra}

The vector fields associated with the Lie symmetries in \eqref{eq:5} read as follows
\begin{equation}
\begin{aligned}
X_{1}&={x}\partial_x+3t\partial_t-\lambda\partial_{\lambda}-u\partial_u-\overline u\partial_{\overline u},\\
X_{2}&=\partial_x,\\
X_{3}&=\partial_t,\\
X_4&=2u\partial_u-2\overline u\partial_{\overline u}+\psi\partial_{\psi}-\chi\partial_{\chi},\\
Y_{\{K(\lambda)\}}&=K(\lambda)\left(\psi\partial_{\psi}+\chi\partial_{\chi}\right),
\end{aligned}
\label{eq:6}
\end{equation}
such that $X_1$ accounts for the usual scaling symmetry, $X_2$ and $X_3$ refer to space and time translations, $X_4$ describes complex phase translations in fields and eigenfunctions, and $Y_{K(\lambda)}$ represents a phase shift in the eigenfunctions. 
Symmetries $X_1-X_4$ extend the original symmetries of the complex mKdV equation \eqref{eq:1} to the spectral problem, whilst the generator $Y_{K(\lambda)}$ is the only symmetry of the Lax pair itself. Nevertheless, this last symmetry is trivially introduced by the linearity of the equations, and it does not provide any further information.

The Lie algebra defined by the vector fields $\langle X_1,X_2,X_3,X_4\rangle$ can be easily addressed by exploring their commutation relations, giving rise to the following nontrivial results,
\begin{equation}
[X_1,X_2]=-X_2,\qquad [X_1,X_3]=-3X_3,
\end{equation}
which define a four-dimensional real Lie algebra which is solvable and decomposable $\langle X_1,X_2,X_3\rangle\oplus X_4$ and can be classified as $A_{3,5}^a\oplus A_1$ for $a=\frac{1}{3}$ \cite{patera1976,patera1977}. 

\section{Similarity reductions}\label{sec:5}

Lie symmetries can be exploited to derive solutions of a given equation in terms of solutions of lower dimensional differential
equations. In particular, if a system of PDEs is invariant under a one-parameter symmetry group, then it is possible to reduce the total number of independent variables by one. This can be achieved by means of the invariants of the system, through the integration of the characteristic system 
\begin{equation}
\frac{dx}{\xi_{x}}=\frac{dt}{\xi_{t}}=\frac{d\lambda}{\xi_{\lambda}}=\frac{du}{\eta_{u}}=\frac{d\overline{u}}{\eta_{\overline{u}}}=\frac{d\psi}{\eta_{\psi}}=\frac{d\chi}{\eta_{\chi}},
\label{eq:7}
\end{equation}
allowing us to introduce the new reduced variables, reduced fields and reduced eigenfunctions as follows
\begin{table}[H]
\centering
\begin{tabular}{lll}
\hline
& Original variables & New reduced variables\\
\hline\hline
Independent variables $\qquad$ & $x, t, \lambda\qquad$ & $z,\,\Lambda$\\[.1cm]
Fields $\qquad$ &$u(x,t), \overline{u}(x,t)\qquad$ & $F(z), \overline{F}(z)$\\[.1cm]

Eigenfunctions $\qquad$ & $\psi(x,t,\lambda),\,\chi(x,t,\lambda)\qquad$ & $ \Psi(z,\Lambda),\,\Phi(z,\Lambda)$\\
\hline
\end{tabular}
\label{table:1}
\end{table}
The only symmetries that yield nontrivial reductions are those related to the transformations of the independent variables, \textit{i.e.} the ones associated with $k_1,\,k_2$ and $k_3$. The remaining symmetries provide trivial reductions, since they are related to phase shifts over the fields and eigenfunctions that are satisfied due to the complexity and linearity of the Lax pair. Then, it is possible to set $k_4=0,\,K(\lambda)=0$ without loss of generality. In the following, we  study two different cases, corresponding to whether the constant $k_1$ is taken as nonzero or null, respectively.

\subsection{$k_1\neq 0$}
The integration of the characteristic system \eqref{eq:7} in the most general case provides the following results:
\begin{itemize}
\item Reduced variable and reduced spectral parameter
\begin{equation}
z=\frac{k_1x+k_2}{k_1^{\frac{2}{3}}\left(3k_1t+k_3\right)^{\frac{1}{3}}},\qquad\qquad\qquad\qquad \Lambda=k_1^{-\frac{1}{3}}\left(3k_1t+k_3\right)^{\frac{1}{3}}\lambda.
\end{equation}
\item Reduced fields
\begin{equation}
u(x,t)=k_1^{\frac{1}{3}}\left(3k_1t+k_3\right)^{-\frac{1}{3}}F(z),\qquad\quad \overline{u}(x,t)=k_1^{\frac{1}{3}}\left(3k_1t+k_3\right)^{-\frac{1}{3}}\overline{F}(z). 
\end{equation}
\item Reduced eigenfunctions
\begin{equation}
\psi(x,t,\lambda)=\Psi(z,\Lambda),\qquad\qquad\qquad \qquad\chi(x,t,\lambda)=\Phi(z,\Lambda).
\end{equation}
\item Reduced spectral problem
%
\begin{subequations}
\begin{equation}
\begin{aligned}
\Psi_{z}&=-\frac{1}{2}\Lambda\Psi-F\Phi,\\
\Phi_{z}&=\overline F\Psi+\frac{1}{2}\Lambda\Phi    
\end{aligned}    
\end{equation}
\vspace{-0.5cm}
\begin{equation}
\begin{aligned}
\Lambda\Psi_{\Lambda}&=\left(\frac{\Lambda\left(\Lambda^2-z\right)}{2}+\Lambda F\overline F-\overline F F_z+F\overline F_z\right)\Psi+\left(\Lambda^2 F-\Lambda F_z+2 \overline F F^2 +F_{zz}-zF\right)\Phi,\\
\Lambda\Phi_{\Lambda}&=-\left(\frac{\Lambda\left(\Lambda^2-z\right)}{2}+\Lambda F\overline F-\overline F F_z+F\overline F_z\right)\Phi-\left(\Lambda^2 \overline F+\Lambda \overline F_z+2 \overline F^2 F +\overline F_{zz}-z\overline F\right)\Psi,    
\end{aligned}    
\end{equation}
\end{subequations}
If the reduced parameter $\Lambda$ has a nontrivial expression (depending on $x$ or $t$), it can be proven that the original spectral problem in $1+1$ dimensions reduces to a reduced Lax pair also in $1+1$ dimensions, where now the independent variables are $z$ and $\Lambda$. Nevertheless, the compatibility condition of such reduced spectral problem retrieves a reduced ordinary differential equation in terms of $z$, as it is illustrated below. 
\item Reduced equation
\begin{equation}
\begin{aligned}
F_{zzz}+6F\overline F F_z\,-zF_z\,-F&=0,\\
\overline F_{zzz}+6\overline F F \overline F_z-z\overline F_z-\overline F&=0.
\end{aligned}    
\end{equation}

For the particular case $F=\overline F$, this system becomes after integration
\begin{equation*}
F_{zz}+2F^3-zF+a=0,
\end{equation*}
which is essentially the second Painlevé transcendent $\text{P}_{\text{II}}$ \cite{ac91,painleve} with arbitrary parameter $a$.
\end{itemize}

\subsection{$k_1=0,\,k_2\neq 0,\,k_3\neq 0$}
An analogous procedure allows us to compute the group invariants for this case from \eqref{eq:7}, yielding
\begin{itemize}
\item Reduced variable and reduced spectral parameter
\begin{equation}
z=\sqrt{\frac{k_2}{k_3}}\left(x-\frac{k_2}{k_3}t\right),\qquad\qquad\qquad \Lambda=\sqrt{\frac{k_3}{k_2}}\,\lambda.
\end{equation}
\item Reduced fields
\begin{equation}
u(x,t)=\sqrt{\frac{k_2}{k_3}}\,F(z),\qquad\qquad\quad \overline{u}(x,t)=\sqrt{\frac{k_2}{k_3}}\,\overline{F}(z)   .
\end{equation}
\item Reduced eigenfunctions
\begin{equation}
\psi(x,t,\lambda)=\Psi(z,\Lambda),\qquad\qquad\quad \chi(x,t,\lambda)=\Phi(z,\Lambda) .  
\end{equation}
\item Reduced spectral problem
%
%
\begin{subequations}\label{eq:17}
\begin{equation}\label{eq:17a}
\begin{aligned}
&\Psi_{z}=-\frac{1}{2}\Lambda\Psi-F\Phi,\\
&\Phi_{z}=\overline F\Psi+\frac{1}{2}\Lambda\Phi,\\
\end{aligned}  
\end{equation}  
\vspace{-0.6cm}
\begin{equation}\label{eq:17b}
\begin{aligned}
&\left(\frac{\Lambda\left(1-\Lambda^2\right)}{2}-\Lambda F\overline F+\overline F F_z-F\overline F_z\right)\Psi-\left(\Lambda^2 F-\Lambda F_z+2 \overline F F^2 +F_{zz}-F\right)\Phi=0,\\
&\left(\frac{\Lambda\left(1-\Lambda^2\right)}{2}-\Lambda F\overline F+\overline F F_z-F\overline F_z\right)\Phi+\left(\Lambda^2 \overline F+\Lambda \overline F_z+2 \overline F^2 F +\overline F_{zz}-\overline F\right)\Psi=0,  
\end{aligned}
\end{equation}
\end{subequations}
\item Reduced equation
\begin{equation}
\begin{aligned}
F_{zzz}+6F\overline F F_z\,-F_z&=0,\\
\overline F_{zzz}+6\overline F F \overline F_z-\overline F_z&=0.
\end{aligned}    
\label{eq:18}
\end{equation}

Notice that for the real case $F=\overline F$, equation (\ref{eq:18}) can be integrated twice as
\begin{equation*}
F_z^2+F^4-F^2=aF+b,
\end{equation*}
with $a,b$ arbitrary constants. The ODE above can be solved through the elliptic integral \cite{byrd}
\begin{equation*}
\int\frac{dF}{\sqrt{aF+b+F^2-F^4}}=\int dz.
\end{equation*}
\end{itemize}

The spectral problem \eqref{eq:17} does not constitute a proper Lax pair for the reduced equations.  Nevertheless \eqref{eq:17b} is a linear system for $\Psi$ and $\Phi$ whose compatibility therefore requires
\begin{equation}
\begin{aligned}
&\left(\frac{\Lambda\left(1-\Lambda^2\right)}{2}-\Lambda F\overline F+\overline F F_z-F\overline F_z\right)^2+\\&+\left(\Lambda^2 F-\Lambda F_z+2 \overline F F^2 +F_{zz}-F\right)\left(\Lambda^2 \overline F+\Lambda \overline F_z+2 \overline F^2 F +\overline F_{zz}-\overline F\right)=0, 
\end{aligned}\label{eq:19}
\end{equation}
where the reduced spectral parameter $\Lambda$ plays the role of an implicit first integral of the system \eqref{eq:18}. It is straightforward to check that, with the aid of (\ref{eq:18}), the derivative of (\ref{eq:19}) is zero.

\section{Conclusions}\label{sec:6}

The first part of the article focuses on the application of the singular manifold method to the complex version of the mKdV equation defined in \eqref{eq1}. This technique allows us to obtain the associated spectral problem through the relation among the singular manifold and the eigenfunctions. Moreover, the systematic application of the singular manifold method to the spectral problem itself leads to the identification of binary Darboux transformations. These transformations can be used to derive an iterative procedure to obtain solutions. Soliton solutions and rogue waves are presented and thoroughly analyzed in this article.

In the second part of the article an alternative approach based on Lie symmetry analysis has been performed. The standard Lie method, when applied directly to the spectral problem, provides both the symmetry transformations for the eigenfunctions, spectral parameter and the fields. In this case, the set of classical Lie symmetries depends on four arbitrary real constants and a single arbitrary function of the spectral parameter $\lambda$. This procedure generalizes the concept of Lie symmetries to the spectral problem, which are fully consistent, as expected, with the Lie symmetries for the original equation. The commutation relations and the classification of the resulting Lie algebra have been thoroughly studied. Finally, two different nontrivial similarity reductions for the associated spectral problem have been obtained. In the first case, the symmetry reduction method allows us to derive the associated spectral problem for the reduced equation, which turns out to be an ordinary 
nonautonomous
differential equation, where the reduced spectral problem plays the role of a new independent variable for the reduced eigenfunctions. In the second case, the similarity reduction directly retrieves an implicit first integral of the analyzed system.

\subsection*{Acknowledgements}

This  research has been supported by MICINN (Grant PID2019-106820RB-C22) and
Junta de Castilla y Le\'on (Grant SA121P20). P. Albares also acknowledges  a postdoctoral contract ``Margarita Salas para la formación de jóvenes doctores'' granted by the Spanish Ministerio de Universidades and Universidad de Salamanca.

\label{lastpage}
\end{document}